\documentclass[]{aa}
\usepackage{graphicx}
\usepackage{natbib}
\usepackage{txfonts}

\newcommand{\sub}[1]{_{\rm #1}}

\newcommand{\OI}{[O\,{\sc i}]}

\newcommand{\CI}{[C\,{\sc i}]}

\newcommand{\changed}{}

\newcommand{\thirdchanged}{}
\newcommand{\fourthchanged}{}

\begin{document}
\title{Spectroscopic \CI{} mapping of the infrared dark cloud G48.65-0.29}
\author{V.~Ossenkopf\inst{1,2,3} \and 
C.~W.~Ormel\inst{2,3,4} \and 
R.~Simon\inst{1} \and K.~Sun\inst{1}  \and 
J.~Stutzki\inst{1}}
\institute{I. Physikalisches Institut der Universit\"at 
zu K\"oln, Z\"ulpicher Stra\ss{}e 77, 50937 K\"oln, Germany
\and
SRON Netherlands Institute for Space Research, P.O. Box 800, 9700 AV 
Groningen, Netherlands
\and
 Kapteyn Astronomical Institute, University of Groningen, P.O. Box 800, 
 9700 AV Groningen, Netherlands
\and
Max-Planck-Institut 
f{\"u}r Astronomie, K{\"o}nigstuhl 17, 69117 Heidelberg, Germany
}

\date{Received: July 16, 2008; accepted October 7, 2010}

\abstract
{We report the first spectroscopic mapping of an atomic carbon line 
  in an infrared dark cloud (IRDC). 
}
{By observing the spatial distribution of the \CI{} emission in an
  IRDC, comparing it with the $^{13}$CO emission and the known
  distribution of internal heating sources, we can quantify the role
  of internal and external UV irradiation in the production of atomic
  carbon.  
}
{We used the $2 \times 4$ pixel SMART receiver of the KOSMA
  observatory on Gornergrat to map the \CI{} $^3P_1 - {}^3P_0$ line in
  the IRDC G48.65-0.29 and compared the resulting spectra with data
  from the BU-FCRAO $^{13}$CO 1--0 Galactic Ring Survey.
}
{The \CI{}/$^{13}$CO effective beam temperature ratio {\thirdchanged falls
  at about 0.3 with local deviations by less than a factor two.\rm} 
  All velocity components seen in $^{13}$CO are also detected in
  \CI{}. We find, however, significant differences in the morphology
  of the brightest regions seen in the two tracers. While $^{13}$CO
  basically follows the column density distribution derived from the
  near-infrared (NIR) extinction and the submm continuum, the \CI{} emission
  peaks at the locations of the two known NIR point
  sources. We find C\,{\sc i}/CO abundance ratios {\thirdchanged between 0.07 
  and 0.13\rm}
  matching the lower end of the range previously measured in
  star-forming regions.
}
{Evaluating the relative importance of the irradiation by embedded
  sources and by the Galactic interstellar UV field, we find that
  in G48.65-0.29 most \CI{} emission can be attributed to externally illuminated
  surfaces.
  Embedded sources have a significant impact on the overall 
  abundance distribution of atomic carbon as soon as they
  are found in an evolved state with noticeable NIR flux.
}

\maketitle

\section{Introduction}

Chemical models of interstellar gas predict that atomic carbon, C{\sc i},
occurs only in a transitional layer characterised by a moderate UV field
which allows ionised carbon to recombine, but still leads to the
dissociation of CO \citep{LeBourlot, Oka2004, Roellig2006}. {\thirdchanged
In the picture of a simple cloud geometry with a well-defined 
interior and exterior of a cloud,} \CI{} is thus
supposed to trace surfaces of molecular clouds exposed to UV
radiation either from the external interstellar radiation field or
from embedded stars \citep{MolSpa}. Its abundance and emission
is modelled in the framework of photon-dominated regions (PDRs).
{\thirdchanged Molecular cloud observations show, however, that
their structure is rather described by a fractal distribution, where
UV radiation can deeply penetrate and surfaces are found throughout
the whole cloud \citep{Falgarone2004},} {\fourthchanged leading to
a strong correlation in the distribution of atomic carbon and
molecular tracers.}

Bright \CI{} emission has been observed towards several star-forming
regions \citep[e.g.,][]{Plume, IC348}, where the intense UV radiation
from young stars creates many PDRs throughout the clumpy ambient
molecular clouds. To understand the origin of the \CI{} emission and
the Galactic abundance distribution of atomic carbon it is, however, 
essential to distinguish
the mutual importance of PDRs within molecular clouds, created by star
formation, and the PDRs at their outer surfaces produced by the
interstellar radiation field illuminating them
\citep{DraineBertoldi}. \cite{Pineda} detected \CI{} towards 
B68, only seeing the 
standard interstellar radiation field, but didn't find a 
PDR model that fitted all of the observed lines.

To help resolving the question on the relative importance of inner
and outer PDRs for the production of atomic carbon we decided to map 
the \CI{} emission from
an infrared dark cloud (IRDC), a site of potentially massive star
formation in a very early stage \citep{Rathborne, Simon}. IRDCs were
detected as dark patches in front of the bright diffuse IR
emission from the Milky Way mapped by ISO and the MSX satellite
\citep{Perault, Egan}. Their column densities exceed $10^{23}$
cm$^{-2}$, producing the strong dust absorption that led to their
discovery. IRDC masses range from $10^3$ to $10^4$ $M\sub{\sun}$, the
gas is typically cold ($T < 25$~K) and dense \citep[$n> 10^5$
  cm$^{-3}$,][]{Simon}. Recent follow-up observations have revealed
signs of active star (cluster) formation in numerous IRDCs 
\citep{Rathborne, Rathborne2}.

We have observed the IRDC G48.65-0.29 \citep{Simon}, a cloud at a
distance of 2.5~kpc in the direction towards W51, that was previously
studied in several molecular transitions by \citet{Shipman}, in the
450~$\mu$m and 850~$\mu$m continuum by \citet{Ormel}, and in the
mid-infrared (MIR), using the Spitzer Space Telescope, by Kraemer et
al. \citep[proposal 3121, see][]{Wiel}. These observations have
revealed several submm condensations and IR point sources
indicating deeply embedded protostars.

This configuration is an ideal test bed to distinguish the role of
internal and external {\fourthchanged UV fields} for the 
distribution of atomic carbon,
because the protostars are still too deeply embedded to allow their UV
radiation to illuminate remote parts of the cloud. We find a clear
distinction between inner and outer surfaces. By observing the spatial
distribution of the \CI{} emission and correlating this with the
distribution of molecular material on the one hand and with the
distribution of young stars within the IRDC on the other hand, we can
discriminate between the effects of internal and external {\fourthchanged 
radiation} for the formation of atomic carbon.

In Sect. 2 we describe the observations and the supplementary data
used in the analysis. Section 3 summarises the results and in Sect. 4,
we derive abundance ratios for a comparison with reference data and
give a first interpretation of the results. Section 5 summarises the
conclusions of the paper.

\section{Observations}

We have used the Sub-Millimeter Array Receiver for Two frequencies
\citep[SMART,][]{SMART} at the K\"olner Observatorium f\"ur
Sub-Millimeter Astronomie \citep[KOSMA,][]{KOSMA} 3~m telescope on
Gorner\-grat, Switzer\-land, to observe the G48 IRDC in the 492~GHz \CI{}
$^3P_1 - {}^3P_0$ line. The observations were taken in March 2004 and
February 2005. At that time, the SMART receiver was equipped with 4
pixels for the 460-490~GHz atmospheric window and 4 pixels for the
810~GHz window. The double sideband system temperature at 492~GHz was
about 250~K for three of the pixels, the fourth pixel was
under-pumped, so that we ignore all data from that pixel here.

We have mapped the cloud in a double beam-switch (DBS) raster mode using
a chopper throw of 6 arcminutes. {\changed Although the \CI{} emission is
in general extended, we can exclude any significant self-chopping
effects due to the spatial and velocity structure of the IRDC as
shown in the Appendix.}
The HPBW of the KOSMA telescope at 492~GHz is 55$''$ and the main 
beam efficiency $\eta\sub{mb}=62$\,\%.
The pointing accuracy of the telescope is
better than 10$''$, we typically measured a 1~$\sigma$ pointing scatter
of about 3$''$ in pointing checks at various sources.
The SMART pixel separation is 115$''$ and we have added in the mapping
two telescope pointings to fill the pixel spacing so that we obtained 
a slightly undersampled map with a spacing of 38$''$.

The IF signal was analysed with $4\times 1$~GHz acousto-optical
spectrometers with a spectral resolution of 1.5~MHz, corresponding to
a velocity resolution of 0.91~km/s at 492~GHz. The channel spacing
is 1.04~MHz, corresponding to 0.63~km/s. The high-frequency channel of
SMART was tuned to the upper (809~GHz) \CI{}$^3P_2 - {}^3P_1$ line, but due
to the higher receiver noise ($T\sub{sys}({\rm DSB}) \approx 650~K$) and the 
lower atmospheric transparency at
that frequency this provided only an upper limit of about 5~K for 
the upper \CI{} line.

For the analysis, we compare the new \CI{} data data with the
$^{13}$CO 1--0 maps taken in the Boston University -- Five College
Radio Astronomy Observatory (BU-FCRAO) Galactic Ring Survey
\citep{FCRAO}. They are given at a velocity resolution of
0.21~km/s. While regridding the FCRAO data from Galactic to equatorial
coordinates, in which the \CI{} map was taken, the data were resampled
to the same spatial resolution of 55$''$.  We provide all spectra on
the original antenna temperature scale. 

{\changed As both \CI{} and $^{13}$CO show emission extending beyond the
main beams of the telescopes, the measured intensity is given by the
superposition of the main beam detection and some error beam pickup.
In the Appendix we have estimated the effective beam efficiencies for
this configuration using the known main and error beam efficiencies
and the approximate configuration of the IRDC. To convert the
antenna temperature to this effective beam temperature scale,
the KOSMA data have to be divided by an effective beam efficiency
of 0.69 and the FCRAO data by 0.70. Since these numbers are almost
identical, all ratios are unaffected by the beam coupling, so that they
have the same value when taken on the antenna temperature or the 
effective beam temperature scale.}

\section{Results}

\begin{figure*}
\includegraphics[angle=90,width=16.8cm]{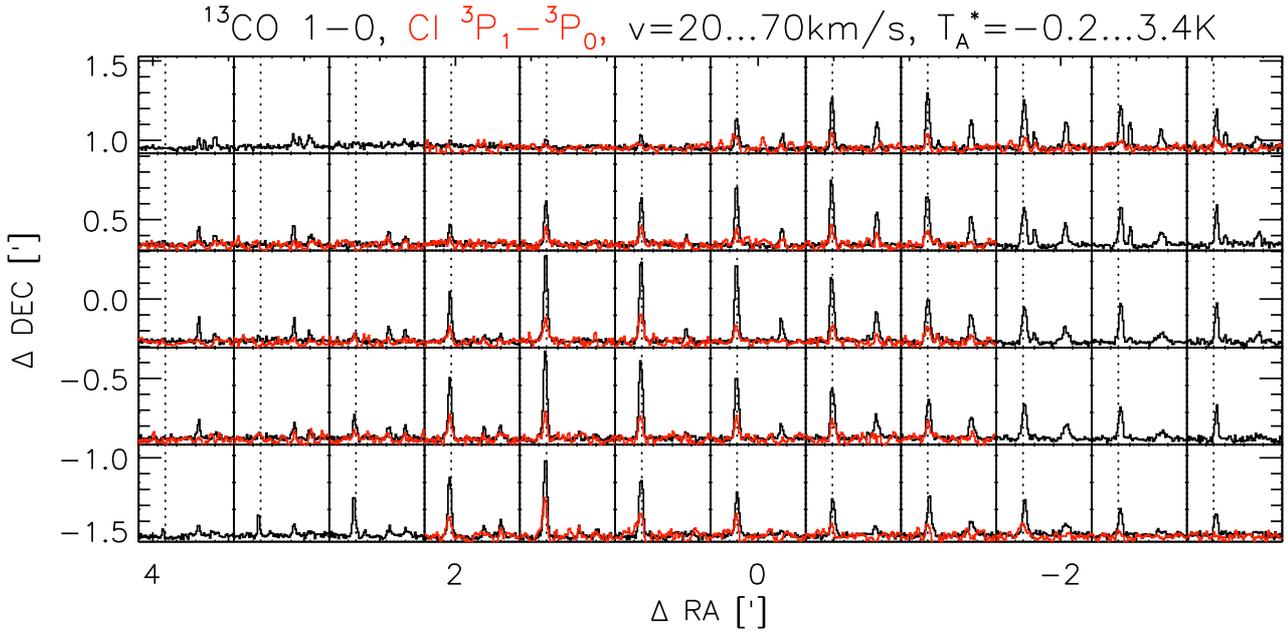}
\caption{Line profiles measured on $^{13}$CO 1-0 (black, BU-FCRAO
  Galactic Ring Survey) and in \CI{} $^3P_1 - {}^3P_0$ (red, KOSMA)
  towards the IRDC G48.65-0.29. All positions are given relative to
  the zero position of $19^h21^m44.7^s$, $13^{\circ}49'31.7''$
  (J2000). For each point we show the spectrum covering the velocity
  range from 20 to 70~km s$^{-1}$ and an antenna temperature range from -0.2
  to 3.4~K.{\thirdchanged The positions without \CI{} spectra were
  either not covered in the mapping or covered by a bad pixel
  providing a noise level much higher than the line intensity.
  The dotted lines at 34~km~s$^{-1}$ mark the LSR velocity of the IRDC.}
}
\label{fig_1}
\end{figure*}

\begin{figure}
\includegraphics[angle=90,width=\columnwidth]{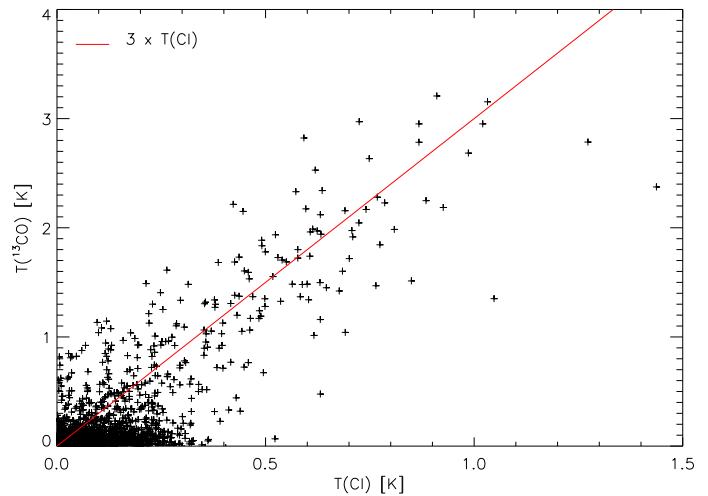}
\caption{Scatter plot of the measured $^{13}$CO and \CI{} intensities
for different channels and points across the cloud. The red line indicates
a \CI{}/$^{13}$CO intensity ratio of 1/3.}
\label{fig_scatter}
\end{figure}

\begin{figure*}
\includegraphics[angle=90,width=16.8cm]{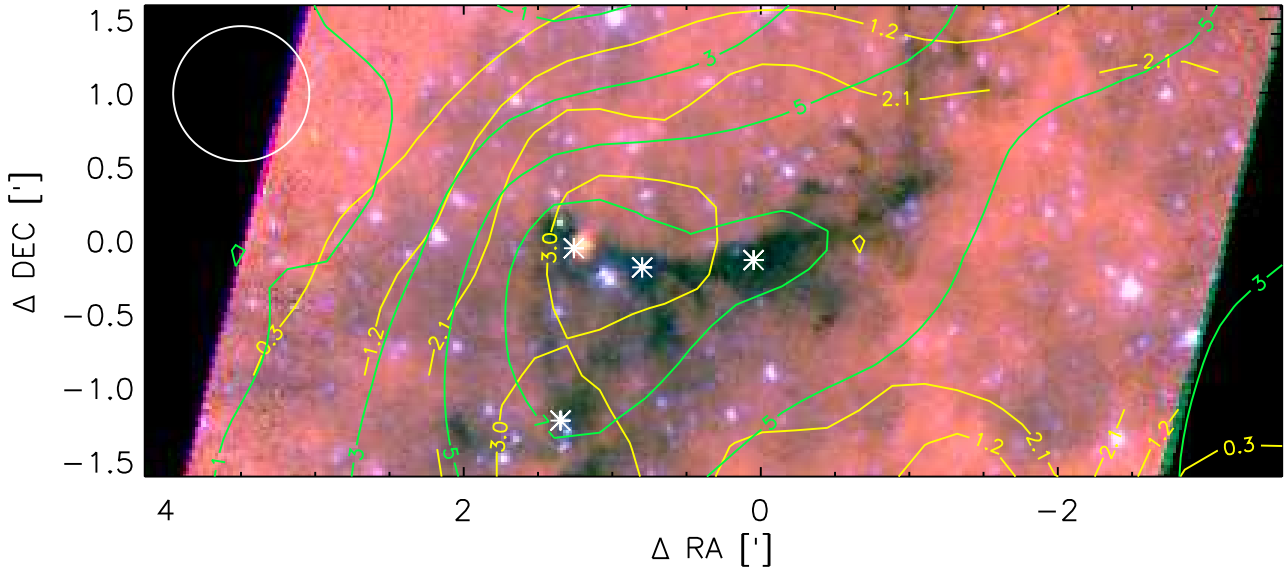}
\caption{Integrated intensity contours in $^{13}$CO 1-0 (green) and 
in \CI{} $^3P_1 - {}^3P_0$ (yellow) on top of a Spitzer IRAC false-colour
image, composed of logarithmic intensities in the 8$\mu$m
channel (red), the 5.8$\mu$m channel (green), and the 3.6$\mu$m
channel (blue). The contours are labelled with intensities [K km s$^{-1}$]
integrated from 30 to 41~km s$^{-1}$. The positions of the
submm cores {\thirdchanged  P1, P2, and EP (from west to east) 
identified by \citet{Ormel} and the southern 24~$\mu$m
source S2 from \citet{Wiel} are indicated by asterisks. The circle 
indicates the beam size of the \CI{} observations, the $^{13}$CO data
are smoothed to the same resolution.}}
\label{fig_spitzer}
\end{figure*}

Figure \ref{fig_1} shows all measured \CI{} spectra on top of the
corresponding FCRAO $^{13}$CO 1--0 spectra covering the G48 IRDC.  We
find {\changed the same spectral features in $^{13}$CO and
\CI{}. There are four dominant velocity components, at 34~km/s,
40~km/s, 52~km/s, and 57-59~km/s, but also some emission at 
intermediate velocities between the components at 34 and 40~km/s and 
the components at 52 and 58~km/s, indicating bridges of gas between 
the components.  Within the noise limits we find no significant
differences in the spectral shapes of the \CI{} and $^{13}$CO 1--0 
lines.}  \citet{Simon} identified
the 34~km/s component with the IRDC based on the morphological
similarity of the $^{13}$CO component and the mid-IR dark cloud in the
MSX image. The 52~km/s component has a kinematic distance of more than
4~kpc representing a separate cloud which appears accidentally in the
same line-of-sight but is unrelated to the IRDC.

{\thirdchanged Figure \ref{fig_scatter} shows a scatter plot of the measured
 \CI{} and $^{13}$CO 1--0 intensities. The typical \CI{}/$^{13}$CO intensity 
ratio is 1/3, the vast majority of the points can be described by ratios
between 0.2 and 0.6. We find no systematic deviations of the ratio as
function of channel velocity, i.e. the same ratio holds in the line
centre and in the line wings and for most of the velocity components.
The component at 52~km/s shows some more values} at the upper end of the
range while the component at 58~km/s tends towards the lower ratios. Both 
components are not related to the IRDC as they fall at different kinematic
distances.  {\thirdchanged At the velocities of the IRDC,} we find 
{\changed relatively small variations in the ratio} between the \CI{} and
the $^{13}$CO intensities for all velocities and all positions where
the lines are detected. \CI{} and $^{13}$CO show the same spatial extent,
but there are some significant differences in the
morphology of the emission. In particular, there is remarkably strong
\CI{} emission in the southern part of the cloud, where the $^{13}$CO
emission is weaker than in the cloud centre.

In Fig. \ref{fig_spitzer}, we show an overlay of the integrated
intensities of the 34~km/s line in $^{13}$CO and \CI{} on top of the
recent Spitzer image of the cloud \citep{Wiel}. One can easily
recognise the good correlation between the MIR extinction and
the $^{13}$CO emission when taking the different angular resolution
into account. The spatial distribution of the \CI{} emission
deviates from this structure, showing two intensity peaks with
one of them located south of the main IRDC structure at the map 
boundary.

\section{Interpretation}

A full analysis of all the spatial and velocity structure visible in
Fig. \ref{fig_1} goes beyond the scope of this paper. We provide
only a first analysis, restricted to the 34~km/s component associated
with the IRDC and start with some morphological considerations.

\citet{Ormel} identified three submm cores in the IRDC as possible
sites of early massive star formation. They are located at offset
positions of 1.26$'$,-0.05$'$ (P1), 0.80$'$,-0.18$'$ (P2), and
0.05$'$,-0.13$'$ (EP) marked by asterisks in Fig. \ref{fig_spitzer} 
\citep[see also][]{Wiel}. These cores basically
line-up along the horizontal filament in Fig. \ref{fig_spitzer} and
are also traced by the $^{13}$CO emission. \citet{Wiel} identified 
24~$\mu$m point sources in all three submm cores (P1 containing two
sources), but also an additional strong 24~$\mu$m source at
1.35$'$,-1.21$'$ (their S2, visible as blue point source in 
Fig. \ref{fig_spitzer}), south of P1, a position where no submm
core was seen. This southern part of the cloud, traced by the IR
extinction and the $^{13}$CO emission, is visible in the
850~$\mu$m continuum, but shows no pronounced condensations.
The point sources in the western cores (P2 and EP) have not been
detected in the IRAC bands while the point sources in the eastern core
(P1) and the southern part show a significant flux even at 3.6~$\mu$m
with the southern source also matching a 2MASS point source. This
indicates that the latter are either in a more evolved state or at a
somewhat lower column density relative to the observer.

Comparing this complementary information with the \CI{} emission shows
that \CI{} neither traces the column density of the cloud nor is
concentrated at the surfaces of the cloud, but that it is peaked
at the sites of intense NIR radiation. If we assume that the large-scale
structure of the IRDC, in particular the surface to volume ratio,
is similar in all parts of the cloud, the internal, more evolved 
{\fourthchanged radiation} sources remain as the main cause
for the enhancement of atomic carbon. The very deeply embedded sources
present in the western cores do not contribute to a local enhancement
of atomic carbon, but as
soon as the sources become visible in the NIR, they change
the C\,{\sc i} abundance.  Nevertheless, we have to emphasise that we
observe relatively small changes of less than a factor two in the
\CI{} intensity.  The global C\,{\sc i} distribution is still
dominated by the overall structure of the IRDC reflecting the external
{\fourthchanged illumination}.

\begin{table}
\caption{Parameters derived at four example positions in the
cloud}\label{tab_parms}
\begin{center}
\begin{tabular}{ll|rrrr}
\hline
Points\tablefootmark{a} & & EP & P1 & S2 & average\tablefootmark{b}\\
Offsets & & 0$'$,0$'$ & 1.3$'$,0$'$ & 1.3$'$,-1.3$'$ & \\
\hline
$\int T\sub{b}($\CI{}$) dv$ & (K km s$^{-1}$) & 3.1 & 4.4 & 5.0 & 1.6 \\
$\int T\sub{b}(^{13}{\rm CO}) dv$ & (K km s$^{-1}$) & 10.4 & 11.7 & 9.3 & 5.1\\
$T[$C{\sc i}$]/T[{\rm ^{13}CO}]\hspace*{-0.4cm}\mbox{}$ & & 0.29 & 0.37 & 0.53 & 0.31\\
\hline
$N[$C{\sc i}$]$\tablefootmark{c} & ($10^{16}$ cm$^{-2}$) & 5.8 & 8.4 & 9.4 & 3.0 \\
$N[^{13}{\rm CO}]$ & ($10^{16}$ cm$^{-2}$) & 1.2 & 1.4 & 1.1 & 0.6 \\
$N[{\rm CO}]$ & ($10^{17}$ cm$^{-2}$) & 7.9 & 8.8 & 7.0 & 3.9 \\
$N[$H$_2]$ & ($10^{22}$ cm$^{-2}$) & 1.0 & 1.1 & 0.9 & 0.5 \\
$N[$C{\sc i}$]/N[{\rm CO}]$ & & 0.07 & 0.09 & 0.13 & 0.08\\
\hline
\end{tabular}
\end{center}
\tablefoot{
\tablefoottext{a}{Within the beam of the \CI{} observations, 
the emission from core P2 
overlaps with that from P1 and EP so that it is not separately listed.}
\tablefoottext{b}{Spatial average over the \CI{} map.}
\tablefoottext{c}{\fourthchanged Using the geometric mean of 
emissivities at $T\sub{ex}=10~$K and $25$~K.}
}
\end{table}

We can use the available complementary information on the structure
of the IRDC and on the physics of PDRs to get a rough estimate for
the temperature of the cloud. In spite of some differences, 
all PDR models predict that the bulk of atomic carbon is
found at temperatures between {\thirdchanged about} 10~K and 25~K 
for sources with low or intermediate radiation fields
\citep{LeBourlot,Roellig2007}\footnote{\thirdchanged Some
uncertainties exist with respect to the temperature of the hottest
few percent of atomic carbon, in particular in regions with
high radiation fields. There, large differences exist between
different PDR model predictions and this is also the range where the
steady-state assumption used by most models is least justified.
A parameter scan using the KOSMA-$\tau$ PDR model showed that
higher average temperatures of atomic carbon can occur for
$n/ M\times \chi  \ga 10^6$~cm$^{-3}/M_{\sun} \chi_0$, i.e.
for small dense clumps in a radiation field exceeding
the average interstellar radiation field $\chi_0$. The known
properties of the IRDC cores are sufficiently far apart from
that boundary so that the uncertainties do not affect our
estimate.}.
Using these values we can 
deduce the C\,{\sc i} column density from a single line applying 
an LTE analysis. The LTE emissivity of the [C\,{\sc i}] $^3P_1 - {}^3P_0$ 
transition varies between 
$3.7 \times 10^{-17}$ and $7.6 \times 10^{-17}$ K~km/s~cm$^{2}$ for
temperatures between 10 and 25~K 
{\fourthchanged ($6.2 \times 10^{-17}$ K~km/s~cm$^{2}$ at 15~K)} 
and the critical density is about $10^3$ cm$^{-3}$. {\fourthchanged
To minimise the relative error of the column density due to the 
temperature uncertainty we use the geometric mean of the two extremes,
i.e. $5.3 \times 10^{-17}$ K~km/s~cm$^{2}$, and obtain} 
a column density estimate that should be accurate within a
factor 1.4.  

Table \ref{tab_parms} summarises the results for four sets of
{\changed effective} beam temperatures, measured at the locations of the western
core (EP), the eastern core (P1), the position of the southern 
24~$\mu$m source (S2), and the map average. {\changed The observed range of 
[C\,{\sc i}] brightness temperatures} is reflected by atomic carbon column 
densities of $3 - 9 \times 10^{16}$ cm$^{-2}$. The corresponding 
beam-integrated optical depths are at most 0.25, so that we can
neglect self-absorption effects.

Using a similar approach for the $^{13}$CO column densities is
somewhat less accurate because the molecule exists over a much larger
temperature range. Here, we can rely on the modelling of the
G48.65-0.29 IRDC by \citet{Ormel}.  They found that only a very small
part of the submm cores is heated to temperatures above 20~K while the
majority of the gas is subject to temperatures between 10 and 15~K,
{\fourthchanged i.e. in a range at the lower end of the typical
temperatures} of atomic carbon. {\fourthchanged This is somewhat
counter-intuitive, as one would usually expect $^{13}$CO deeper
in the cloud than atomic carbon at locations better shielded from 
the UV radiation and consequently also significantly colder. 
However, higher densities in the inner cloud also provide a 
more efficient gas-grain interaction that can heat the gas again,
even allowing for non-monotonous temperature distributions.}
The LTE emissivity of the $^{13}$CO 1-0 transition is $7.4 \times 10^{-16}$
K~km/s~cm$^{2}$ at 20~K and $1.0 \times 10^{-15}$ K~km/s~cm$^{2}$ at 10~K
{\fourthchanged ($8.7 \times 10^{-15}$ K~km/s~cm$^{2}$ at 15~K)},
the critical density is well below $10^3$ {\fourthchanged cm$^{-3}$. 
Using again a mean conversion factor, i.e.} 
$8.6 \times 10^{-16}$ K~km/s~cm$^{2}$ should be accurate within 
20\,\%.  Assuming a $^{12}$CO/$^{13}$CO abundance ratio of 65 
\citep{Langer}, we can translate this into CO column densities and
C\,{\sc i}/CO abundance ratios. Applying the standard CO/H$_2$ 
abundance ratio of $8 \times 10^{-5}$ from \citet{Blake} gives also a
rough estimate for the total column densities. They are consistent 
with the numbers obtained by \citet{Ormel} and \citet{Simon}. 
All numbers are summarised in Table \ref{tab_parms}.

We find C\,{\sc i}/CO abundance ratios between {\changed 0.07 and 0.13}. 
The C\,{\sc i}/CO abundance ratio previously measured in star-forming 
regions covers the range between about 0.03 and 3\footnote{Lower 
values have been reported only towards the DR15 H{\sc ii} region by \citet{Oka}.}
\citep{Bhaswati, Little, IC348}. The values found for G48.65-0.29
{are found at the lower end} of this range and this is consistent with the picture of atomic
carbon being produced by star-formation processes in which more
quiescent clouds, {\fourthchanged without internal UV fields,} 
show lower C\,{\sc i}/CO ratios than more evolved, active star-forming sites.

The LTE analysis applied above ignores, however, the possible spatial
separation of the regions dominating the [C\,{\sc i}] and the
$^{13}$CO 1-0 emission. A physically self-consistent model of
the cloud can be obtained using the known density structure of the 
submm cores obtained by \citet{Ormel}. The roughly spherical core 
structure is well approximated by the KOSMA-$\tau$ PDR model 
\citep{Roellig2006}. {\changed This model simulates a}
{\thirdchanged spherical cloud with a radial density profile 
given by a $r^{-1.5}$ decay covering a factor five in radii 
around a core with constant density. Assuming external
illumination by an isotropic UV radiation field and cosmic rays,
the model computes the stationary chemical and temperature
structure by solving the coupled detailed balance of heating,
line and continuum cooling and the chemical network using the
UMIST data base of reaction rates \citep{UMIST} expanded by separate
entries for the $^{13}$C chemistry \citep[see][for details]{Roellig2007}. 
The model configuration provides a good match to the quiescent cores
in the IRDC and may reflect the situation for most of the material
in the IRDC. Here, atomic carbon is produced in a layer at the
surface of the dense material and in the more extended diffuse 
material. In the inner core the material is predominantly 
molecular. 

This approach is, however, not appropriate for the cores 
with internal {\fourthchanged radiation sources}, 
P1 and S2, where we expect some additional 
atomic carbon production in the centres. Consequently, the 
KOSMA-$\tau$ PDR model should be able to reproduce the
\CI{} $^3P_1 - {}^3P_0/ ^{13}$CO 1-0 intensity ratio of about 0.3 
found at the extinction peak and in the cloud average, but
will only provide a lower limit to the \CI{}/$^{13}$CO ratio
for the two warm cores. When comparing the line ratios
at the different positions (see Tab. \ref{tab_parms}),
we find that} the internal {\fourthchanged radiation} from 
P1 produces only a slight 
enhancement of the atomic carbon abundance over the value produced
by the external UV illumination while for the southern source
{\fourthchanged internal and external illumination} provide
similar contributions.

\begin{figure}
\includegraphics[angle=90,width=8.8cm]{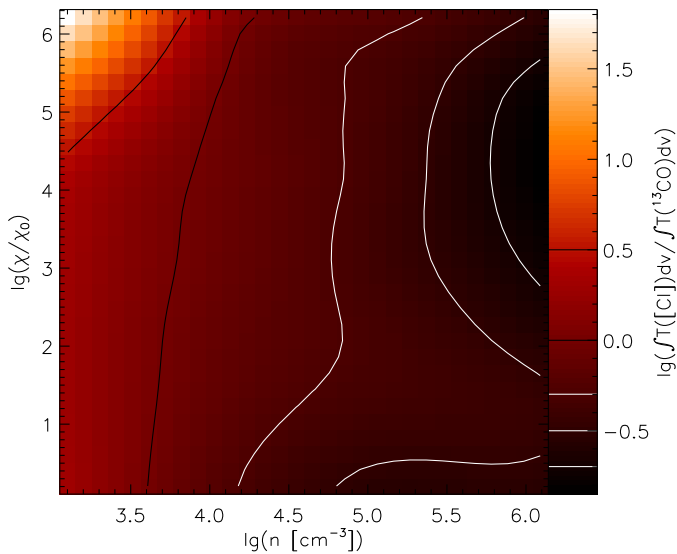}
\caption{\CI{} $^3P_1 - {}^3P_0/ ^{13}$CO 1-0 line intensity
ratio predicted by the KOSMA-$\tau$ PDR model for a spherical,
centrally-condensed, externally illuminated cloud core of 
$100~M\sub{\sun}$ as a function of the surface density and the
strength of the UV field. {\thirdchanged Contours are drawn
at ratios of 0.2, 0.32, 0.5, 1.0, and 3.2 to emphasise 
the region of interest.}}
\label{fig_kosmatau}
\end{figure}

Figure \ref{fig_kosmatau} shows the predicted intensity ratios
{\thirdchanged as a function of cloud surface density and radiation 
field} provided by the PDR extraction 
tool\footnote{http://hera.ph1.uni-koeln.de/$\sim$pdr/}
for the known core mass of about $100~M\sub{\sun}$ 
{\thirdchanged and solar metallicity\footnote{The web tool provides integrated
line intensities in terms of photon energies, i.e. units of
erg~s$^{-1}$~cm$^{-2}$~sr$^{-1}$. For intensity units of K~km~s$^{-1}$,
the conversion factor of $2k\nu^3/c^3 = 1.025 \times 10^{-15} (\nu/{\rm GHz})^3$
has to be applied. This gives 
a conversion factor for the $\int T([$C{\sc I}$])dv/\int T(^{13}$CO$)dv$
intensity ratio of 0.0112.}.
White contours indicate the area of ratios matching the observed
range. The ratio of about 0.3, i.e. 
lg$\big \{\int T([$C{\sc I}$])dv/\int T(^{13}$CO$)dv\big \} \approx -0.5$,
observed within most of the cloud and marked by the central
white contour is only observed at high densities. For
radiation fields in the order of the standard interstellar radiation
field $\chi_0$ \citep{DraineBertoldi}, a model with a clump surface 
density of about $3\times 10^4$~cm$^{-3}$ reproduces the 
observed ratio. The model predicts average temperatures of
atomic carbon and $^{13}$CO of 12~K and 9~K, respectively, 
consistent with the estimates used above for the LTE analysis,
but somewhat at the low edge as the model does not contain internal
{\fourthchanged UV radiation} sources.

When comparing the model with the results from the fit of the dust continuum and HCO$^+$ 
emission by \citet{Ormel}, we find a major discrepancy in the gas density.
\citet{Ormel} derived surface densities of only about 
$10^3$~cm$^{-3}$. When taking internal clumpiness of the cores 
into account, local densities grow by a factor ten (core P1)
or five (P2 and EP), but are still significantly lower
than the model density fitting the \CI{}/$^{13}$CO ratio.
For the lower densities, the KOSMA-$\tau$ model always predicts 
significantly higher line ratios. Models} that assume a further
break-up of the cores into smaller clumps have a higher surface to
volume ratio, thus predicting even higher \CI{}/$^{13}$CO ratios.
{\thirdchanged Lower ratios could be achieved for radiation fields significantly
below the standard field $\chi_0$ but that would be hard to explain.
Excitation from W51 can be excluded due to the large 
relative distance \citep[][]{Reid} so that we expect
fields in the order of $\chi_0$. 
A match could also be achieved when assuming unrealistically
high metallicities $Z>3$. With realistic parameters and the
known masses and densities, the model
predicts ratios above 0.5, only observed at the peaks with internal 
{\fourthchanged radiation sources} where the model is expected to 
underestimate the ratio by neglecting
this {\fourthchanged contribution}. Invoking other model geometries
than spherical clumps can only worsen the situation as they will
always increase the surface to volume ratio of the cloud. 
{\fourthchanged Fractionation effects will hardly resolve the
discrepancy as the KOSMA-$\tau$ PDR model includes the full
$^{13}$C chemistry. But we have to admit that reaction
rates for isotope-selective reactions are quite uncertain at
low temperatures (see R\"ollig et al., in preparation).}
Thus, there is a clear contradiction between the measured 
\CI{}/$^{13}$CO line ratios and the model predicting higher values.}

The same contradiction
was noticed by \citet{Bensch} for the MCLD 123.5+24.9 translucent
cloud. They suggested a pre-shielding by dust in a thin ionised gas
as a possible solution. Time-dependent chemistry has been shown
to tend towards the opposite direction, increasing the C\,{\sc i}
abundance \citep{Stoerzer1997, Oka2004}, but a fully time-dependent model,
including turbulent mixing, still awaits development.

\section{Conclusions}

Our observations of \CI{} in an IRDC showed
an overall very good correlation between atomic carbon and
molecular gas, seen in $^{13}$CO 1-0, both in the spatial and
the velocity structure. This is naturally explained by a fractal
cloud structure illuminated by the external Galactic UV field
\citep{MolSpa}. The quantitative \CI{} $^3P_1 - {}^3P_0/^{13}$CO 1-0
ratios, however, cannot be explained by available steady-state 
models. A spherical PDR clump model predicts higher line
ratios than observed. In a fractal structure with more internal
surfaces, this ratio should even grow further. More sophisticated
models are needed. {\changed Future observations of the emission
of atomic oxygen at 4.7~THz may help to resolve the puzzle as
\OI{} is another, independent tracer of {\fourthchanged illuminated,} dense cloud surfaces.}

With the significant enhancement of the \CI{} intensity 
observed towards the southern source, visible in the Spitzer
IRAC bands, we can provide a first distinction between
the roles of inner and outer PDRs if we assume that the outer
structure is similar in all parts of the IRDC. A fraction of the \CI{}
emission comparable to that from the outer cloud surfaces, 
exposed to the interstellar radiation field,
can be attributed to the internal {\fourthchanged radiation} source.
This means that
any young stellar object that has created a sufficient cavity in 
the parental cloud, turning visible in the NIR, produces
a considerable fraction of the total \CI{} emission.
For most of the star-forming regions observed so far in \CI{}, the
emission will thus stem from inner surfaces.

\begin{acknowledgements}
We thank an anonymous referee for many helpful comments.
This work has been supported by the Deut\-sche
For\-schungs\-ge\-mein\-schaft through grant 494B. 
It has made use of NASA's Astrophysics Data System Abstract Service.
\end{acknowledgements}

\appendix
\section{Effects of extended emission}
\label{appx_a}

\begin{figure}
\includegraphics[angle=0,width=\columnwidth]{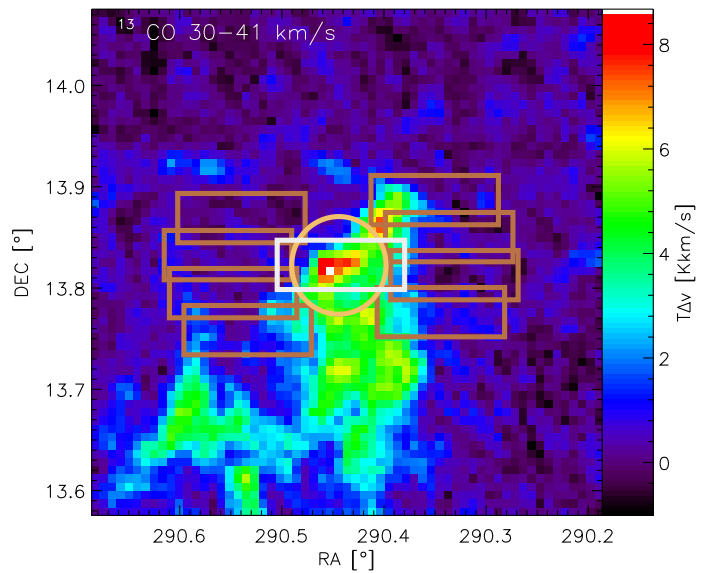}
\caption{Map of the large-scale emission as measured by the $^{13}$CO
  1--0 line. The map covered by \CI{} observations
  is shown as gray rectangle in the centre, the brown rectangles
  around that map indicate the locations of the OFF maps
  taken in the double beam-switch scheme at the different times
  and dates of the observations. The circle indicates the size
  of the KOSMA error beam at the \CI{} frequency. The FCRAO
  error beam at the $^{13}$CO 1--0 frequency is smaller by a factor
  three and is not shown here.}
\label{fig_appx}
\end{figure}

Both $^{13}$CO and \CI{} are known to show very extended emission
so that it needs to be investigated to which degree the measured
intensity at each point may be affected by this emission. Two
effects have to be considered: self-chopping and intensity pick-up
by the extended error beam.
Taken the similarity of the emission of $^{13}$CO and \CI{}
we can estimate these effects from the larger $^{13}$CO 1--0
FCRAO map showing also the environment of the IRDC. It is displayed
in Fig. \ref{fig_appx} using the same integration range as in Fig.
\ref{fig_spitzer}.

The \CI{} observations were taken in double beam-switch mode
using a fixed chop throw of 6$'$ in azimuthal direction. The sky
rotation relative to the azimuthal direction then leads to variable
OFF positions depending on date and time of the observation.
In Fig. \ref{fig_appx} we have indicated the area of OFF maps
covered in the course of the observation runs. We see that the
centre of the map is never affected by self-chopping effects,
but that the north-west corner and the east edge of the map
could be affected by self chopping. As we have not detected any
significant \CI{} emission at these points, this might only
mean that \CI{} is slightly more extended than shown in Fig.
\ref{fig_spitzer}.


We have to admit, however, that the spectra show some self-chopping 
effects at frequencies out of the considered velocity interval.
This can be seen as negative features around 44~km/s in some
spectra in the south-west part of the map and around 55~km/s
in the south of the map.
These contributions
stem from gas components south-east of the IRDC. As all these 
features only show up outside of the velocity integration interval
assigned to the IRDC, self chopping effects can be neglected
for the integrated \CI{} intensities.


The extended emission will also lead to additional intensity
pick-up in the error beams of the telescopes. The error beam of the
KOSMA telescope, produced by a slight misalignment of the individual
panels has a size of about 330$''$ \citep{Kramer}.
Comparing the moon efficiency of 80\,\% to the main beam efficiency
of 62\,\% provides a maximum error beam efficiency of 18\,\%.
We have estimated the error beam pickup by convolving the FCRAO
map with a 330$''$ Gaussian, weighted by the error beam efficiency,
and comparing it to the map values weighted by the main beam
efficiency. We find that at the emission peaks, the error beam
contribution falls between 8 and 12\,\%, over most of the map it amounts
to about 15\,\% but it reaches a value of 30\,\% at the south-west
corner of the map. Consequently, the true \CI{} emission
might be somewhat less extended in the south-west direction than shown
in Fig. \ref{fig_spitzer}. Due to the error-beam pickup we 
overestimate the local emission when dividing the measured intensity 
by the main-beam efficiency only. Using an effective beam efficiency
of 69\,\% instead of 62\,\% gives a correction for this effect that
should be accurate to about 10\,\% over most of the observed map.

The error beam of the FCRAO telescope at the $^{13}$CO 1--0 frequency
has a width of about 100$''$. The main beam efficiency is about
48\,\% and the measured moon efficiency 72\,\% \citep{Ladd}. Repeating the
deconvolution exercise for this smaller error beam gives a relatively
constant ratio of about 0.45 for most of the map which is only
slightly smaller than the ratio between error beam and main beam 
efficiency. Consequently, we will use a value of 70\,\%, only
slightly below the moon efficiency as the effective beam efficiency 
to calibrate the $^{13}$CO. 

This approach may slightly underestimate the intensity as the moon
efficiency is only an upper limit for the sum of main-beam and 
error-beam efficiency and some small components at the scale of a few
arcminutes might contribute, but the residual error from this
contribution will certainly fall below 10\,\%.


\begin{thebibliography}{}
%
\bibitem[Blake et al.(1987)]{Blake}
Blake, G.A.; Sutton, E.C.; Masson, C.R.; Phillips, T.G. 1987, ApJ 315, 621

\bibitem[Bensch et al.(2003)]{Bensch}
Bensch, F.; Leuenhagen, U.; Stutzki, J.; Schieder, R. 2003, ApJ 591, 1013

\bibitem[Draine \& Bertoldi(1996)]{DraineBertoldi}
Draine, B.T., Bertoldi, F. 1996, ApJ 468, 269

\bibitem[Egan et al.(1998)]{Egan} Egan, M. P.; Shipman, R. F.; Price, S. D.;
Carey, S. J.; Clark, F. O.; Cohen, M. 1998, ApJ 494, L199

\bibitem[Falgarone et al.(2004)]{Falgarone2004}
Falgarone, E.; Hily-Blant, P. \& Levrier F. 2004, ApSS 292, 89

\bibitem[Graf et al.(2003)]{SMART}
Graf, U.U.; Heyminck, S.; Michael, E.A.; Stanko, S.; Honingh, C. E. et al.
2003, SPIE, 4855, 32

\bibitem[Jackson et al.(2006)]{FCRAO}
Jackson, J. M.; Rathborne, J. M.; Shah, R. Y.; 
Simon, R.; Bania, T. M.;  et al. 2006, ApJS 163, 145

\bibitem[Langer \& Penzias(1990)]{Langer}
Langer, W.D.; Penzias, A.A. 1990, ApJ 357, 477

\bibitem[Le Bourlot(1993)]{LeBourlot}
Le Bourlot, J.; Pineau Des Forets, G.; Roueff, E.; Flower, D. R. 1993,
A\&A 267, 233

\bibitem[Kramer et al.(2004)]{Kramer}
Kramer, C.; Jakob, H.; Mookerjea, B.; Schneider, N.; Brüll, M.; Stutzki, J.
2004, A\&A 424, 887

\bibitem[Ladd \& Covey(2000)]{Ladd}
Ladd, E. F.; Covey, K. R. 2000, ApJ 536, 380

\bibitem[Little et al.(1994)]{Little}
Little, L.T.; Gibb, A.G.; Heaton, B.D.; Ellison, B.N.; Claude S.M.X.
1994, MNRAS 271, 649

\bibitem[Mookerjea et al.(2006)]{Bhaswati}
Mookerjea, B.; Kramer, C.; R\"ollig, M.; Masur, M. 2006, A\&A 456, 235

\bibitem[Oka et al.(2001)]{Oka}
Oka, T.; Yamamoto, S.; Iwata, M. et al. 2001, ApJ 558, 176

\bibitem[Oka et al.(2004)]{Oka2004}
Oka, T.; Iwata, M.; Maezawa, H. et al. 2004, ApJ 602, 803

\bibitem[Ormel et al.(2005)]{Ormel}
Ormel, C. W.; Shipman, R. F.; Ossenkopf, V.; Helmich, F. P. 2005,
A\&A,439, 613

\bibitem[Ossenkopf et al.(2007)]{MolSpa}
Ossenkopf, V.; R\"ollig, M.; Cubick, M.; Stutzki, J. 2007, in 
{\it Molecules in Space and Laboratory}, by J.L. Lemaire \& F. Combes (eds.),
S. Diana publ., p.95


\bibitem[P\'erault et al.(1996)]{Perault}
P\'erault, M.; Omont, A.; Simon, G. 1996, A\&A 315, L165

\bibitem[Pineda \& Bensch(2007)]{Pineda}
Pineda, J. L.; Bensch, F. 2007, A\&A 470, 615

\bibitem[Plume et al.(2000)]{Plume} 
Plume, R.; Bensch, F.; Howe, J.E.; Ashby, M.L.N.; Bergin, E.A. et al. 
2000, ApJ 539, L133

\bibitem[Rathborne et al.(2006)]{Rathborne}
Rathborne, J.M.; Jackson, J.M.; Simon R. 2006,  ApJ 641, 389

\bibitem[Rathborne et al.(2007)]{Rathborne2}
Rathborne, J.M.; Simon R. 2006; Jackson, J.M. 2007,  ApJ 662, 1082

\bibitem[Reid(1993)]{Reid}
Reid, M.J. 1993, ARAA 31, 345
 
\bibitem[R\"ollig et al.(2006)]{Roellig2006}
R\"ollig, M.; Ossenkopf, V.; Jeyakumar, S.; Stutzki, J.; Sternberg, A.
2006, A\&A 451, 917

\bibitem[R\"ollig et al.(2007)]{Roellig2007}
R\"ollig, M.;Abel, N. P.; Bell, T.; Bensch, F.; Black, J. et al.
2007, A\&A 467, 187

\bibitem[Shipman et al.(2003)]{Shipman}
Shipman, R.F.; Frieswijk, W.; Helmich, F.P. 2003, in
{\it Galactic Star Formation Across the Stellar Mass Spectrum}, by 
J.M. de Buizer \& N.S. van der Bliek (eds.), p. 252

\bibitem[Simon et al.(2006)]{Simon}
Simon, R.; Rathborne, J. M.; Shah, R. Y.; Jackson, J. M.; Chambers, E. T. 2006,
ApJ 653, 1325


\bibitem[St\"orzer et al.(1997)]{Stoerzer1997}
St\"orzer, H.; Stutzki, J.; Sternberg, A. 1996, A\&A 323, L13

\bibitem[Sun et al.(2008)]{IC348}
Sun, K.; Ossenkopf, V.; Mookerjea, B.; R\"ollig, M.; Kramer, C.;
Cubick, M.; Stutzki, J. 2008, A\&A, 489, 207

\bibitem[Van der Wiel \& Shipman(2008)]{Wiel}
van der Wiel, M.H.D.; Shipman, R.F. 2008, A\&A, 490, 655

\bibitem[Winnewisser et al.(1986)]{KOSMA}
Winnewisser, G.; Bester, M.; Ewald, R. 1986, A\&A 167, 207

\bibitem[Woodall et al.(2007)]{UMIST}
Woodall, J.; Ag{\'u}ndez, M.; Markwick-Kemper, A.~J.; Millar, T.~J. 
2007, A\&A 466, 1197
%
\end{thebibliography}
\end{document}